\documentclass[12pt,preprint]{aastex}

\usepackage{graphicx}        
\usepackage{natbib}         
\usepackage{amssymb}        
\usepackage{amsmath}
\usepackage{color}           
\usepackage{url}             
\usepackage{epstopdf}

\shorttitle{Excitation of gravity waves by fingering convection, and the formation of compositional staircases in stellar interiors}
\shortauthors{Garaud et al.}

\newcommand{\be}{{\bf e}}
\newcommand{\bk}{{\bf k}}
\newcommand{\bu}{{\bf u}}
\newcommand{\bF}{{\bf F}}
\newcommand{\bR}{{\bf R}}
\renewcommand{\Pr}{{\rm Pr}}
\newcommand{\Nu}{{\rm Nu}}
\newcommand{\bnabla}{{\mathbf{\nabla}}}

\begin{document}


\title{Excitation of gravity waves by fingering convection, and the formation of compositional staircases in stellar interiors}


\author{P. Garaud$^1$, M. Medrano$^1$, J. M. Brown$^2$, C. Mankovich$^2$ \& K. Moore$^{1,2,3}$}
\affil{$^1$ Department of Applied Mathematics and Statistics, Baskin School of Engineering, University of California Santa Cruz, 1156 High Street, Santa Cruz CA 95060  \\
$^2$ Department of Astronomy and Astrophysics, University of California Santa Cruz, Interdisciplinary Sciences Building, Santa Cruz, CA 95064, USA \\
$^3$ TASC, University of California Santa Cruz, Santa Cruz CA 95060}


\begin{abstract}
Fingering convection (or thermohaline convection) is a weak yet important kind of mixing that occurs in stably-stratified stellar radiation zones in the presence of an inverse mean-molecular-weight gradient. Brown et al. (2013) recently proposed a new model for mixing by fingering convection, which contains no free parameter, and was found to fit the results of direct numerical simulations in almost all cases. Notably, however, they found that mixing was substantially enhanced above their predicted values in the few cases where large-scale gravity waves, followed by thermo-compositional layering, grew spontaneously from the fingering convection. This effect is well-known in the oceanographic context, and is attributed to the excitation of the so-called ``collective instability''. In this work, we build on the results of Brown et al. (2013) and of Traxler et al. (2011b) to determine the conditions under which the collective instability may be expected. We find that it is only relevant in stellar regions which have a relatively large Prandtl number (the ratio of the kinematic viscosity to the thermal diffusivity),  $O(10^{-3})$ or larger. This implies that the collective instability cannot occur in main sequence stars, where the Prandtl number is always much smaller than this (except in the outer layers of surface convection zones where fingering is irrelevant anyway). It could in principle be excited in regions of high electron degeneracy, during He core flash, or in the interiors of white dwarfs. We discuss the implications of our findings for these objects, both from a theoretical and from an observational point of view. 
\end{abstract}





\section{Introduction}

Discovered in the late 1950s in the oceanographic context \citep{stern1960sfa}, fingering -- or ``thermohaline'' -- convection was first introduced to the astrophysical community by \citet{Ulrich1972} and later revisited by \citet{kippenhahn80}. The fingering instability occurs in stars within radiation zones that have an unstable mean molecular weight gradient ($\mu$-gradient hereafter). This situation is often found as a result of material accretion onto a star by anything from a single or multiple planets \citep{vauclair2004mfa,theado12tv}, to material from a dust-enriched or debris accretion disk \citep[e.g.][]{Dealal2013}, or material from a more evolved companion \citep{Ulrich1972,stancliffe2007cem}. It also naturally arises in the vicinity of the H-burning shell in red giant branch (RGB) stars, as discussed by \citet{CharbonnelZahn07} and \citet{Denissenkov2010}, and in thin element-rich layers near the surface of intermediate-mass stars \citep{theado09,Zemskovaal2014}. The fingering instability initially takes the form of  thin tubes, hence the name ``finger'', within which the fluid moves vertically. The tubes rapidly break down, however, as a result of parasitic shear instabilities that develop inbetween them \citep{Radkosmith2012,Brownal2013}, and the fingering instability eventually saturates into a state of homogeneous fingering convection where the typical aspect ratio of the eddies is closer to one \citep{Traxleral2011}.

One of the main challenges in the subject of fingering convection is to model the rate at which it transports or mixes various quantities of interest, such as heat, angular momentum, and chemical species. Until very recently, the only available mixing prescriptions for fingering convection were the ones originally proposed by \citet{Ulrich1972} and \citet{kippenhahn80}. Both were essentially based on dimensional arguments and ad-hoc modeling efforts, that could not, for lack of experimental data, be directly verified. Furthermore, the two estimates for the turbulent diffusion coefficient differed by several orders of magnitude. 

Thanks to advances in supercomputing, however, much progress has recently been made to measure the rate of mixing in fingering convection in numerical experiments, and to develop theoretical models for the collected data. \citet{Traxleral2011} first proposed a simple empirical formula for the turbulent diffusion coefficient for both heat and composition, which was fitted to a substantial number of three-dimensional numerical experiments run at various input parameters (notably, different values of viscosity, thermal diffusivity, compositional diffusivity and background stratifications of temperature and composition). \citet{Brownal2013} later added further simulations and showed that the model of \citet{Traxleral2011}, while perfectly adequate for more strongly stratified systems (ie. systems with a weaker inverse $\mu$-gradient, or systems with stronger thermal stratification), underestimates the mixing efficiency for systems that are very close to being unstable to standard overturning convection. They proposed an alternative model, based this time on first principles, which fits the numerical data remarkably well for most available simulations (in the limit where the compositional diffusivity is smaller than the kinematic viscosity, which is always the case in stellar conditions), and is simple enough to be used in real time in any stellar evolution code.  

Despite this significant progress, much remains to be done to fully characterize the properties of fingering convection in astrophysics. In the related oceanographic context, it is well-known that small-scale homogeneous fingering is itself unstable to secondary large-scale instabilities of various forms including the $\gamma$-instability, which leads to  the formation thermohaline staircases \citep{radko2003mlf}, the collective instability, which excites large-scale gravity waves \citep{stern1969cis,stern2001sfu}, and the intrusive instability, which also gives rise to layering in the presence of horizontal thermo-compositional gradients \citep{holyer83,walsh1995int}. In all three cases, the presence of larger-scale dynamics can significantly enhance the rate of turbulent mixing above that of homogeneous fingering convection \citep[see][for instance]{Brownal2013}, so it is crucial to understand why and when these secondary instabilities arise. 

As shown by \citet{Traxleral2011b}, all three large-scale modes of instability can be studied within the same unifying framework, which consists in deriving so-called ``mean-field'' equations for the large-scale fields while treating the small-scale fingering convection only through its contribution to the turbulent transport of heat, composition and momentum. However, their work was specifically applied to oceanographic conditions; whether similar secondary instabilities also occur in stellar conditions remains the subject of undergoing research. It has recently been shown that the $\gamma$-instability cannot operate in fingering systems in stellar interiors \citep{Traxleral2011,Brownal2013}, but that intrusive instabilities are ubiquitous in the presence of horizontal gradients \citep{medrano2014}. In this paper, we focus on the third form of large-scale instability, namely the collective instability. 

The collective instability was first studied by \citet{stern1969cis} in the oceanographic context \citep[see also][]{stern2001sfu}. As mentioned above, it is a mechanism by which small-scale fingering convection naturally excites large-scale internal gravity waves. By ``large'' we imply waves whose typical wavelength is at least one order of magnitude greater than that of the basic fingering instability. The physical mechanism behind the collective instability is easy to understand, given some basic knowledge about the development of the standard double-diffusive instabilities, which we now briefly summarize (see the review by \citet{Garaud13} for more detail). First, recall that double-diffusive instabilities in general develop in fluids whose overall density is stably stratified, but depends on two different quantities that (1) diffuse at different rates and (2) have opposing contributions to the density stratification. The fingering instability arises whenever the faster-diffusing component (typically, temperature) is stably stratified while the slower-diffusing (typically, composition) component is unstably stratified. The oscillatory instability (which is related to semiconvection), by contrast, develops when the faster-diffusing component is unstably stratified, while the slower-diffusing component is stably stratified \citep{Kato66,bainesgill1969}. This occurs in thermally-unstable regions that are stabilized by a strong $\mu$-gradient. The oscillatory instability takes the form of overstable gravity waves, i.e. oscillatory motions whose amplitude increases exponentially with time.

As clarified by \citet{Radko13}, the collective instability can simply be viewed as a large-scale manifestation of the oscillatory instability, in a system which (on the small scales) is unstable to fingering convection and where the microscopic thermal and compositional diffusivities are therefore augmented by the induced turbulent mixing. Indeed, the key is to note that fingering convection transports composition much more rapidly than it transports heat, so the turbulent diffusivity of the stably stratified field (composition) is much larger than the turbulent diffusivity of the unstably-stratified field (temperature). This situation, as discussed above, gives rise to an oscillatory instability, whose growth rate depends on the fingering fluxes. Furthermore, since the turbulent diffusivities are larger than their microscopic counterparts, the typical lengthscale of the collective modes is also significantly larger than the basic fingering scale. This has advantages and disadvantages for modeling the instability. On the one hand, the separation of scale between the collective instability and the fingering instability allows for the use of mean field theory, as shown by \citet{Traxleral2011b}, a particularly convenient tool in this problem. On the other hand, it also means that the collective instability can only be studied numerically in very large computational domains, which has only recently become feasible. \citet{stellmach2011} were the first to demonstrate the natural emergence of the collective instability in numerical simulations of fingering convection at parameter values similar to those expected in the ocean. They found that the theory developed by \citet{Traxleral2011b} adequately predicts the observed growth rate of the collective modes in that case.

\citet{DenissenkovMerryfield2011} were the first to study the collective instability in the astrophysical context, with particular application to RGB stars. Using the formalism of \citet{Traxleral2011b} together with estimates of the fingering fluxes from 2D numerical simulations, they found that collective modes are not excited. On the other hand, \citet{Brownal2013} saw the emergence of oscillatory features in their 3D direct numerical simulations of fingering convection. This apparent discrepancy can be explained by the difference in the parameters used: the results of \citet{DenissenkovMerryfield2011} were obtained for realistic stellar values of the Prandtl number and diffusivity ratio ($\Pr = \nu/\kappa_T \sim 10^{-6}$ and $\tau = \kappa_\mu/\kappa_T \sim 10^{-7}$ respectively), while the simulations of \citet{Brownal2013} were run, for numerical feasibility, at $\Pr \sim \tau \sim 0.01$ instead. The difference between the findings of \citet{DenissenkovMerryfield2011} and \citet{Brownal2013} raises an interesting question, however: where in parameter space does the collective instability lie, and are there any stars for which it may be relevant? 

The possibility of driving large-scale gravity waves in fingering convection is indeed quite exciting for two reasons. First, and as discussed above, these secondary instabilities are usually accompanied by enhanced transport, and could therefore help solve some of the outstanding problems in stellar evolution associated with ``missing mixing''. Second, the gravity waves themselves may be directly detectable via asteroseismology, should they have sufficiently large spatial structures and amplitudes. Our goal, therefore, is twofold: to perform a comprehensive study of the collective instability and determine under which conditions (i.e. for which values of the microscopic diffusivities, background thermal and compositional stratification) the latter is expected, and then to search across the HR diagram for stellar objects where such conditions may be realized. 

In what follows, we study the collective instability in the more general astrophysical context (i.e. without limiting our study to parameters relevant for RGB stars) in Section \ref{sec:collinst}. We also improve on the study by \citet{DenissenkovMerryfield2011} by using the more recent fingering flux laws of \citet{Brownal2013}, and by comparing our results with large-scale 3D numerical simulations of the collective instability (see Section \ref{sec:num}). As we shall demonstrate the latter is only present for values of the Prandtl number and of the diffusivity ratio down to about $10^{-3}$, in regions of parameter space that are very close to being overturning-unstable. This finding is therefore consistent both with the results of \citet{DenissenkovMerryfield2011} on the absence of the collective instability in RGB stars and on its presence in the numerical simulations of \citet{Brownal2013}. We then look in Section \ref{sec:HRdiag} across the HR diagram and attempt to identify stars which have ``large'' Prandtl number and diffusivity ratio. Our findings are discussed in Section \ref{sec:ccl}.

\section{The collective instability}
\label{sec:collinst}

To model the collective instability, we begin with the standard (non-dimensional) equations of fluid dynamics applied to the problem of fingering convection \citep{radko2003mlf}:
\begin{eqnarray}
\frac{1}{\Pr} \left(\frac{\partial\bu}{\partial t} + \bu \cdot \bnabla \bu \right) & = & -\bnabla p + (T - \mu)\be_z+ \nabla^2 \bu\label{eq:momentum}, \\
\nabla \cdot \bu &=& 0 \label{eq:continuity},  \\
\frac{\partial T}{\partial t} +  \bu \cdot \bnabla T + w  & = & \nabla^2 T \label{eq:heat}, \\
\frac{\partial \mu}{\partial t} +  \bu \cdot \bnabla \mu + \frac{1}{R_0}w  & = & \tau \nabla^2 \mu
\label{eq:composition},
\end{eqnarray}
where $\bu = (u,v,w)$ is the velocity field, and $p$, $T$ and $\mu$ are the pressure, temperature and mean molecular weight perturbations away from the chosen background.
Assuming that the vertical size of the domain under consideration is small compared with the respective local scale heights, the background temperature and mean molecular weight profiles are approximated by linear functions of the vertical coordinate $z$, with constant (dimensional) gradients $T_{0z}$ and $\mu_{0z}$ respectively. 
To arrive at the non-dimensional equations (\ref{eq:momentum})-(\ref{eq:composition}), we have used the following system of units. The unit length is 
\begin{equation}
[l] = d = \left(\frac{\kappa_T\nu}{g\alpha (T_{0z}-T_{0z}^{\rm ad})}\right)^{1/4} =  \left(\frac{\kappa_T\nu}{N_T^2}\right)^{1/4} \, ,
\end{equation}
where $\kappa_T$ is the thermal diffusivity, $\nu$ is the viscosity, $g$ is gravity, $\alpha$ is the coefficient of thermal expansion, $T_{0z}^{\rm ad}$ is the local adiabatic temperature gradient, and $N_T$ is the local buoyancy frequency based on the thermal stratification only.  All of these quantities are also assumed to be constant. The unit time is $[t] = d^2 /\kappa_T$, and the unit velocity is $[v] = \kappa_T/d$. The unit temperature is $[T] = d (T_{0z}-T^{\rm ad}_{0z})$ and finally, the unit mean molecular weight is $[\mu] = \alpha d (T_{0z}-T^{\rm ad}_{0z}) / \beta$, where $\beta$ is the coefficient of ``compositional contraction''. The coefficients $\alpha$ and $\beta$ are thermodynamic derivatives of the equation of state, and given by 
\begin{equation}
\alpha = - \frac{1}{\rho_0} \left(\frac{\partial \rho}{\partial T}\right)_{p,\mu} \mbox{  and } \beta =  \frac{1}{\rho_0} \left(\frac{\partial \rho}{\partial \mu}\right)_{p,T} \, ,
\end{equation} 
where $\rho_0$ is the mean density of the region considered. The three standard non-dimensional parameters of fingering convection thus emerge: the Prandtl number $\Pr = \nu/\kappa_T$, the diffusivity ratio $\tau = \kappa_\mu/\kappa_T$, and the density ratio $R_0 = \alpha (T_{0z}-T^{\rm ad}_{0z})/\beta \mu_{0z}$. A system is linearly unstable to fingering convection when $R_0$ is in the range $[1, \tau^{-1}]$ \citep{stern1960sfa,bainesgill1969}, so we restrict our following analysis to that range only. 

We now filter (\ref{eq:momentum})-(\ref{eq:composition}) to extract the dynamics of scales that are significantly larger than the basic finger scale (which is of the order of $10d$, typically). This leads to  
\begin{eqnarray}
\frac{1}{\Pr} \left(\frac{\partial\overline{\bu}}{\partial t} + \overline{\bu} \cdot \bnabla \overline{\bu} + \bnabla \cdot \bR \right) & = & -\bnabla \overline{p} + (\overline{T} - \overline{\mu})\be_z+ \nabla^2 \overline{\bu} \label{eq:avmomentum}, \\
\nabla \cdot \overline{\bu} &=& 0 \label{eq:avcontinuity},  \\
\frac{\partial \overline{T}}{\partial t} +  \overline{\bu} \cdot \bnabla \overline{T} + \bnabla \cdot \bF_T + \overline{w}  & = & \nabla^2 \overline{T} \label{eq:avheat}, \\
\frac{\partial \overline{\mu}}{\partial t} +  \overline{\bu} \cdot \bnabla \overline{\mu} + \bnabla \cdot \bF_\mu + \frac{1}{R_0} \overline{w}    & = & \tau \nabla^2 \overline{\mu}, \label{eq:avmu}
\end{eqnarray}
where the overbar denotes a local spatial average over the small fingering scales, that we assume commutes with both spatial and temporal derivatives. Defining $\bu' = \bu - \overline{\bu}$ and similarly for $T'$ and $\mu'$, averages of the nonlinear terms in the original equations give rise to the Reynolds stress tensor $\bR = \overline{\bu'\bu'}$, the temperature flux $\bF_T = \overline{\bu' T'}$ and the compositional flux  $\bF_\mu = \overline{\bu' \mu'}$. This system of equations is not closed unless these three quadratic correlation terms are known functions of the large-scale variables $\overline{\bu}$, $\overline{T}$ and $\overline{\mu}$ and of the model parameters $\Pr$, $\tau$ and $R_0$. 

We now follow the procedure proposed by \citet{Traxleral2011b} to model the collective instability. The steps are outlined here for completeness; the reader is referred to the original paper for more pedagogical detail if needed. \citet{Traxleral2011b} proposed the following ``closure'' model: (a) the Reynolds stress is negligible compared with other terms in the momentum equation, (b) the turbulent fingering fluxes are principally vertical, so $\bF_T = F_T \be_z$ and similarly for $\bF_\mu = F_\mu \be_z$. Condition (a) has been verified a posteriori via numerical simulations. Condition (b) on the other hand is only marginally justified as numerical simulations show that the horizontal components of the fluxes can  be up to a third of the vertical component. Nevertheless, we shall make that assumption for simplicity. We now define the thermal Nusselt number $\Nu_T$ and the turbulent flux ratio $\gamma$ as
\begin{eqnarray}
 F_T =  (1-\Nu_T) \left(1 + \frac{\partial \overline{T}}{\partial z} \right) \, ,\nonumber \\ 
 F_\mu = \frac{F_T}{\gamma} \, .
 \label{eq:fluxes}
 \end{eqnarray}
In the final assumption of the closure model, (c), we assume that $\Nu_T$ and $\gamma$ are functions of $\overline{T}$ and $\overline{\mu}$ only through the {\it local} density ratio 
\begin{equation}
R = R_0 \frac{1 + \frac{\partial \overline{T}}{\partial z}}{1 + R_0 \frac{\partial \overline{\mu}}{\partial z}} \, .
\label{eq:localR}
\end{equation}
$\Nu_T$ and $\gamma$ of course also depend on $\Pr$ and $\tau$, so (c) implies $\Nu_T = \Nu_T(R; \Pr, \tau)$ and $\gamma = \gamma(R; \Pr ,\tau)$. Whether assumption (c) is valid or not will be verified a posteriori, if the collective instability model correctly predicts the growth rate of large-scale gravity waves observed in simulations. We do know, however, that (c) is valid in the absence of large-scale temperature and compositional perturbations, since in that case the fluxes can {\it only} depend on the input non-dimensional parameters of the system, $R_0$, $\Pr$ and $\tau$.

Substituting (\ref{eq:fluxes}) and (\ref{eq:localR}) into (\ref{eq:avmomentum})-(\ref{eq:avmu}), we immediately see that there is a trivial solution to the resulting equations: one where all the large-scale fields are zero, namely $\overline{\bu} = \overline{T} = \overline{\mu} = \overline{p} = 0$, where the local density ratio is equal to the background density ratio, $R = R_0$, and where 
$F_T$ and $F_\mu$ are constant and equal to their corresponding values in the absence of large-scale perturbations, namely
\begin{eqnarray}
 F_T =  1-\Nu_T(R_0; \Pr,\tau) \equiv 1 - \Nu_0 \, ,\nonumber \\ 
 F_\mu = \frac{F_T}{\gamma(R_0; \Pr,\tau)} \equiv \frac{F_T}{\gamma_0} \, ,
 \end{eqnarray}
which defines $\Nu_0$ and $\gamma_0$. This solution is merely the state of homogeneous fingering convection one would expect in the absence of large-scale perturbations. 

We now consider that $\overline{T}$, $\overline{\mu}$ and $\overline{\bu}$ are large-scale but small-amplitude perturbations on this homogeneously turbulent state, and linearize the system (\ref{eq:avmomentum})-(\ref{eq:avmu}) around the latter. Linearizing (\ref{eq:localR}) around $R_0$, the local density ratio becomes 
\begin{equation}
R = R_0 \left(1 + \frac{\partial \overline{T}}{\partial z} - R_0 \frac{\partial \overline{\mu}}{\partial z} \right)  \, ,
\end{equation}
and we define the coefficients $A_1 = R_0 (d\gamma^{-1}/dR)_{R=R_0} $ and $A_2 = R_0 (d\Nu_T/dR)_{R=R_0} $ such that
\begin{eqnarray}
\gamma^{-1}(R;\Pr,\tau) \simeq \gamma^{-1}_0 + (R-R_0)  \left(\frac{d \gamma^{-1}}{d R}\right)_{R_0}  \simeq \gamma^{-1}_0 +  A_1 \left( \frac{\partial \overline{T}}{\partial z} - R_0 \frac{\partial \overline{\mu}}{\partial z}\right)  \, , \nonumber \\
\Nu_T(R;\Pr,\tau) \simeq \Nu_0 + (R-R_0) \left(\frac{d \Nu_T}{d R}\right)_{R_0}  \simeq \Nu_0 + A_2 \left( \frac{\partial \overline{T}}{\partial z} - R_0 \frac{\partial \overline{\mu}}{\partial z}\right)  \, .
\end{eqnarray}
Assuming that all quantities can be expressed as linear combinations of normal modes of the kind $\overline{q} = \hat q \exp(ilx + imy + ikz + \Lambda t)$, we then obtain a cubic equation for the growth rate $\Lambda$ of the large-scale modes, i.e. 
\begin{equation}
\Lambda^3 + a_2 \Lambda^2 + a_1 \Lambda + a_0 = 0 \, ,
\label{eq:largecubic}
\end{equation}
where 
\begin{subequations}
\label{full_theory}
\begin{eqnarray}
a_2 & = & |\mathbf{k}|^2(1+\Pr+\tau) + k^2\left[(1-A_1R_0)(\Nu_0-1) + A_2\left(1-\frac{R_0}{\gamma_0}\right) \right], \\
a_1 & = & |\mathbf{k}|^4(\tau\Pr + \tau + \Pr) + k^2|\mathbf{k}|^2 \left[(\tau + \Pr)(A_2 + \Nu_0 - 1) - A_2(1+\Pr)\frac{R_0}{\gamma_0} \right. \nonumber \\
 & &  - A_1R_0(1+\Pr)(\Nu_0-1) \biggr] - k^4A_1R_0(\Nu_0-1)^2 + \Pr\frac{L^2}{|\mathbf{k}|^2}  \left(1-\frac{1}{R_0}\right), \\
a_0 & = & |\mathbf{k}|^6\tau\Pr + k^2|\mathbf{k}|^4\, \Pr\left[(\tau-A_1R_0)(\Nu_0-1) + A_2\left(\tau-\frac{R_0}{\gamma_0}\right)\right] - k^4|\mathbf{k}|^2\Pr R_0A_1(\Nu_0-1)^2 \nonumber \\
 & & + \Pr\frac{L^2}{|\mathbf{k}|^2}\left\{ |\mathbf{k}|^2 \left(\tau - \frac{1}{R_0}\right)  + k^2A_1(1-R_0)(\Nu_0-1) \right. \nonumber \\
 & & \left. - k^2\left[A_2\left(1-R_0\right) + \Nu_0-1\right]  \left(\frac{1}{R_0} - \frac{1}{\gamma_0}\right) \right\}, 
\end{eqnarray}
\end{subequations}
where $|\mathbf{k}|^2 =  l^2 + m^2 + k^2$. This is the same cubic as in \citet{Traxleral2011b}, assuming that there are no background horizontal gradients of temperature and composition. Note that in this limit, there is a perfect symmetry between the $x$ and $y$ directions, and the system dynamics only know about the total horizontal wavenumber $L = \sqrt{l^2 + m^2}$. Also note that the cubic (\ref{eq:largecubic}) has 3 solutions $\Lambda$ for each input $\bk$. In what follows, when referring to ``the growth rate of the collective mode with wave vector $\bk$'' we always implicitly take the root which has the largest real part. 

The only remaining problem is to determine the constants $\Nu_0$, $\gamma_0$, $A_1$ and $A_2$, which are needed to compute $a_0$, $a_1$ and $a_2$. They are determined from the properties of homogeneous fingering convection, and are functions only of the density ratio $R_0$ and of $\Pr$ and $\tau$. While \citet{Traxleral2011b} used values of these coefficients appropriate for fingering convection in an oceanographic context ($\Pr \sim 7$ and $\tau \sim 0.01$), in stellar interiors $\Pr$ and $\tau$ are typically much smaller than one, so their values cannot be used here. However, $\Nu_0$ and $\gamma_0$ can be  directly estimated using the model proposed by \citet{Brownal2013} for mixing by fingering convection:  
\begin{eqnarray}
\Nu_T(R;\Pr,\tau) = 1 + 49 \frac{ \lambda_{\rm fgm}^2 }{l_{\rm fgm}^2 (\lambda_{\rm fgm} + l_{\rm fgm}^2)} \, , \nonumber \\ 
\gamma(R;\Pr,\tau) = R  \frac{ \lambda_{\rm fgm}+\tau l_{\rm fgm}^2 }{\lambda_{\rm fgm} + l_{\rm fgm}^2} \, ,
\end{eqnarray}
for $R = R_0$, where $\lambda_{\rm fgm}$ and $l_{\rm fgm}$ are functions of $R$, $\Pr$ and $\tau$ and are the growth rates and horizontal wavenumbers respectively of the fastest growing modes of the basic fingering instability at these input parameters (recalling that, for the basic instability, the fastest-growing modes have a zero vertical wavenumber). They can be found by solving simultaneously a cubic and a quadratic equation:
\begin{align}
\lambda_{\rm fgm}^3&+b_2\lambda_{\rm fgm}^2+b_1\lambda_{\rm fgm}+b_0=0\textrm{, where} \label{eq:smallcubic} \\
b_2&=l_{\rm fgm}^2(1+\Pr+\tau), \nonumber\\
b_1&=l_{\rm fgm}^4(\tau\Pr+\Pr+\tau)+\Pr\left(1-\frac{1}{R}\right), \nonumber\\
b_0&=l_{\rm fgm}^6\tau\Pr+l_{\rm fgm}^2\Pr\left(\tau-\frac{1}{R}\right) \, , \nonumber
\end{align}
and 
\begin{align}
c_2\lambda_{\rm fgm}^2&+c_1\lambda_{\rm fgm}+c_0=0\textrm{, where} \label{eq:maxk}\\
c_2&=1+\Pr+\tau, \nonumber\\
c_1&=2l_{\rm fgm}^2(\tau\Pr+\tau+\Pr), \nonumber\\
c_0&=3l_{\rm fgm}^4\tau\Pr+\Pr\left(\tau-\frac{1}{R}\right) \, . \nonumber
\end{align}
Note that (\ref{eq:smallcubic}) is simply obtained from (\ref{eq:largecubic}) with $\gamma_0 = \Nu_0 = A_1 = A_2 = 0$ (which recovers the dispersion relation for the basic fingering instability), $R_0 = R$ and $k = 0$, while (\ref{eq:maxk}) is then derived from (\ref{eq:smallcubic}) by maximizing the growth rate $\lambda$ over all possible values of the horizontal wavenumber. The coefficients $A_1$ and $A_2$ are then obtained by differentiating the functions $\Nu_T(R;\Pr,\tau)$ and $\gamma^{-1}(R;\Pr,\tau)$ numerically, as 
\begin{eqnarray}
A_1 = R_0 \frac{\gamma^{-1}(R_0+\epsilon; \Pr,\tau) - \gamma^{-1}(R_0;\Pr,\tau)}{\epsilon} \, ,\nonumber \\
A_2 = R_0 \frac{\Nu_T(R_0+\epsilon; \Pr,\tau) - \Nu_T(R_0;\Pr,\tau)}{\epsilon} \, .
\end{eqnarray}
In practice, we take $\epsilon = 10^{-3} R_0$ in what follows.  

\begin{figure}[h]
\begin{center}
\includegraphics[width=0.9\textwidth]{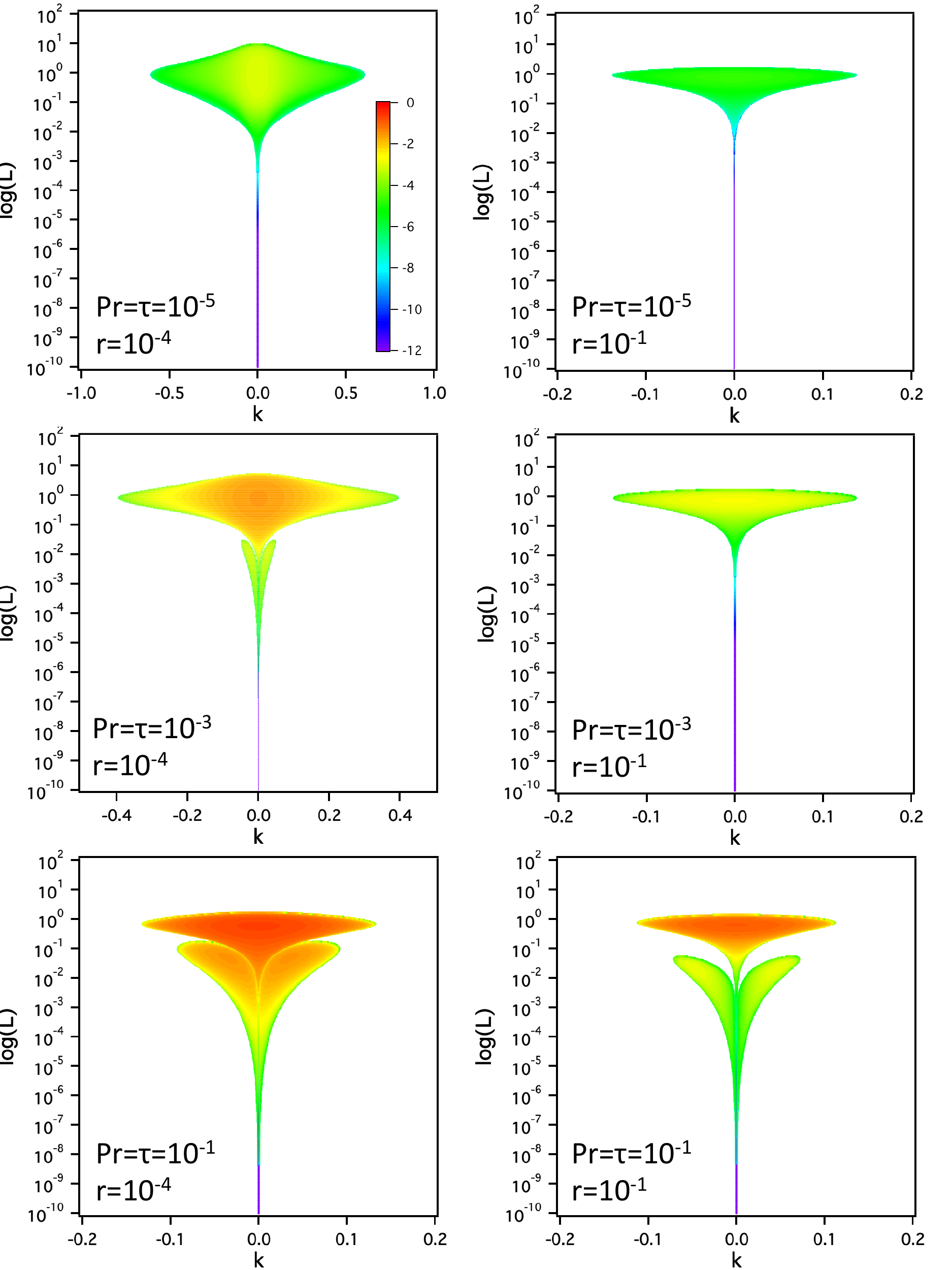}
\caption{Largest value of the real part of $\Lambda$ among the three possible solutions of the cubic (\ref{eq:largecubic}), as a function of vertical wavenumber $k$ (horizontal axis) and horizontal wavenumber $L$ (vertical axis), for various values of the Prandtl number (with $\tau = \Pr$) and reduced density ratio $r = (R_0-1)/(\tau^{-1}-1)$. Areas shown in white have negative growth rates. The ``bulb'' in each plot corresponds to the fingering modes, while the ``leaves'' are the collective instability modes.}
\label{fig:flowerplots}
\end{center}
\end{figure}

Figure \ref{fig:flowerplots} shows the real part of the growth rate $Re(\Lambda)$ of each mode with spatial structure given by a wave vector of horizontal component $L$ and vertical component $k$, for various values of the Prandtl number (assuming $\tau = \Pr$), and for various values of $R_0$. Modes with negative $Re(\Lambda)$ are shown in white. Note that for ease of comparison of the various diagrams, we use the reduced density ratio introduced by \citet{Traxleral2011}, namely 
\begin{equation}
r = \frac{R_0-1}{\tau^{-1} - 1} \, ,
\end{equation}
which is 0 when $R_0 = 1$, i.e. when the system is marginally unstable to standard overturning convection, and is 1 when $R_0 = \tau^{-1}$, ie. when the system is marginally stable. Hence cases with $r = 10^{-4}$ are close to being overturning-unstable, while cases with $r = 0.1$ are only weakly unstable to fingering. 

We immediately note that the growth rate is independent of the sign of $k$, which is expected since the cubic governing $\Lambda$ only depends on $k^2$. The same is true for $L$, so the figure only shows results for positive values of the latter. As found by \citet{Traxleral2011b}, the instability diagram thus drawn takes the shape of a flower. The bulb of the flower centered around $L \sim 1$ and $k = 0$ corresponds to the basic fingering modes, while the leaves, when they exist, are the large-scale gravity-wave modes excited by the collective instability. By contrast with the fingering modes, the collective modes have a complex growth rate, whose imaginary part is the oscillation frequency of the wave. They typically have low $L$ and low (but non-zero) $k$ wave numbers.  Figure \ref{fig:flowerplots} suggests that they only exist for the smaller values of $R_0$ and the larger values of the Prandtl number (and $\tau)$. 

\begin{figure}[h]
\begin{center}
\includegraphics[width=\textwidth]{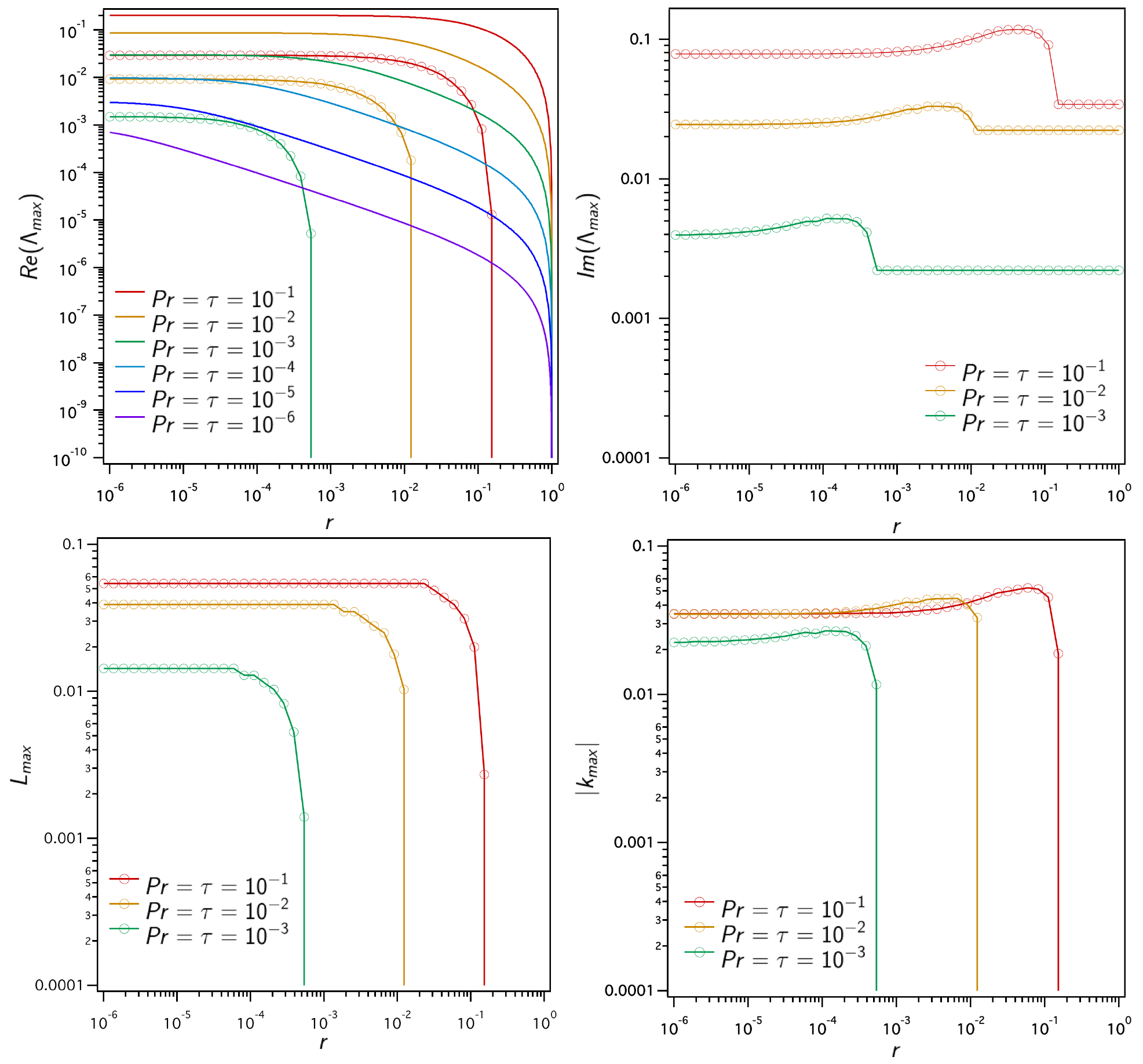}
\caption{Properties of the fastest-growing collective mode (solid lines with circles), in comparison with the properties of the fastest-growing fingering mode (plain solid lines), as a function of the reduced density ratio $r$. Top left: real part of $\Lambda_{\rm max}$, the growth rate of the fastest growing collective mode, and of $\lambda_{\rm fgm}$ (solution of equations (\ref{eq:smallcubic}) and (\ref{eq:maxk})), the growth rate of the fastest growing fingering mode. Top right: Oscillation frequency of the fastest-growing collective instability mode. Bottom left: horizontal wavenumber $L_{\rm max}$ of the fastest growing collective mode.  Bottom right: Vertical wavenumber $k_{\rm max}$ of the fastest-growing collective mode. }
\label{fig:growthrates}
\end{center}
\end{figure}

To see this more clearly, we show in Figure \ref{fig:growthrates} the growth rate (real and imaginary parts) and wavevector of the fastest-growing collective mode as a function of $\Pr$, $\tau$ and of the reduced background density ratio $r$. These are found by maximizing the real part of $\Lambda$ over all modes of wavenumber $L$ and $k$ with non-zero imaginary part.  We see that collective modes exist for $\Pr = \tau = 0.1$, 0.01, and 0.001, but not for $\Pr,\tau \le 10^{-4}$. We also see that for $\Pr, \tau \ge 10^{-3}$, these modes only exist for very low values of $r$ (equivalently, for density ratios very close to one), in other words, for systems whose background stratification is already fairly close to being Ledoux-unstable. Note that this result had already been obtained by \citet{Brownal2013} using the Stern number as a approximate diagnostic of the presence of the collective instability. 

The reason why the collective instability is limited to the higher values of $\Pr$ and $\tau$, and the lower density ratios, is quite simple: this also corresponds to the region of parameter space where the magnitudes of the fingering fluxes (of temperature and composition) are large enough to have a significant effect on the overall heat and compositional transport when compared with pure diffusion. Since the physical mechanism behind the collective instability discussed in Section 1 crucially relies on the turbulent fluxes, it is not surprising to find that the instability disappears when these fluxes are too small. 

Wherever collective modes exist, we find that the real part of $\Lambda$ for the fastest-growing ones is of the order of $\Pr$, which in dimensional terms corresponds to $ \Pr \kappa_T /  d^2 =  \Pr^{1/2} N_T$. For $\Pr \sim 10^{-2}$, this is of the order of $0.1N_T$. The imaginary part of $\Lambda$ is of the same order, so we expect each of these modes to oscillate with a frequency commensurate with their exponential growth rate. Finally, both horizontal and vertical wavenumbers are of the order of a few times $10^{-2}$, so we expect the fastest-growing collective modes wavelengths to be of the order of a few hundred times $d$ or in other words, a few tens of times the width of the fastest-growing fingering mode. What this corresponds to in real dimensional terms depends on the actual dimensional values of the thermal and compositional diffusivities and of the local buoyancy frequency, and will therefore vary from star to star (see Section \ref{sec:ccl} for an example).

\section{Comparison with numerical simulations} 
\label{sec:num}

In order to test the validity of the collective instability theory described above, we now turn to Direct Numerical Simulations (DNS) of fingering convection, and use a computational domain that is sufficiently large to contain the large-scale gravity waves we expect will develop. We use the same code as \citet{Traxleral2011b}, which solves the set of equations (\ref{eq:momentum})-(\ref{eq:composition}) in a triply periodic domain. We present here the results of a simulation at $\Pr = 0.1$ and $\tau = 0.03$, which are the smallest values of the Prandtl number and of the diffusivity ratio we can realistically achieve while at the same time using a computational domain large enough to contain at least a few wavelengths of the fastest-growing collective mode. We use a density ratio of $R_0 = 1.1$, for which the growth of that collective mode is relatively fast. This is the same set of parameters as one of the simulations of \citet{Brownal2013} for which wave-like behavior (and eventual layer formation) was indeed observed. 

At these parameters, the non-dimensional growth rate and horizontal wavenumber $\lambda_{\rm fgm}$ and $l_{\rm fgm}$ of the fastest-growing fingering mode are:
\begin{equation}
\lambda_{\rm fgm} \simeq 0.2  \mbox{   and  } l_{\rm fgm} \simeq 0.75 \, ,
\end{equation}
while the growth rate, horizontal wavenumber and vertical wavenumber of the fastest-growing collective mode are 
\begin{equation}
\Lambda_{\rm coll} \simeq 0.024 \pm 0.10 i \mbox{   ,  } L_{\rm coll} \simeq 0.060  \mbox{   and  } k_{\rm coll} \simeq 0.048 \, .
\end{equation}
As a result, we need a domain whose vertical extent is at least $2\pi / 0.048 \simeq 130d$, with a similar horizontal extent, to contain the fastest-growing mode of the collective instability. For this reason, we cannot use the aforementioned simulation of \citet{Brownal2013}, which was performed in a domain of size $(100d)^3$. In what follows, we use a domain of size $(L_x = 400d,L_y=20d,L_z = 400d)$, which is very thin in the $y$-direction to save on computational time. It is nevertheless thick enough to contain at least two wavelengths of the fastest-growing fingering mode (which is about 8$d$ at these parameters). This was found to be an excellent trade-off that captures the three-dimensional dynamics of the basic fingering instability adequately while being quasi-two-dimensional for the collective instability. The effective resolution in this run is 384 meshpoints per 100$d$. 

\begin{figure}[h]
\begin{center}
\includegraphics[width=\textwidth]{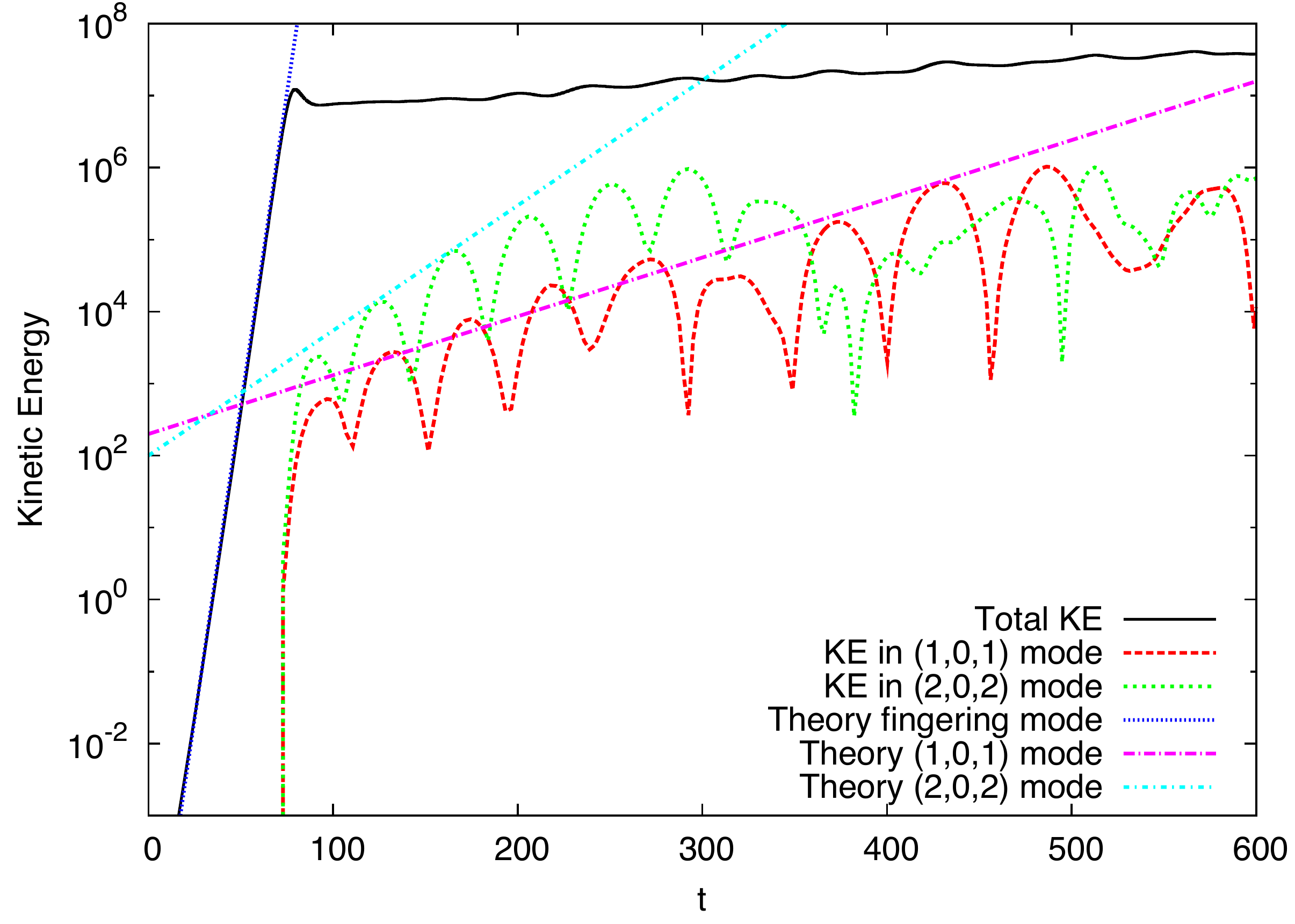}
\caption{Total kinetic energy in the domain, and kinetic energy in selected collective modes. The modes are referenced by the number of wavelengths that fit in the computational domain. Hence the $(1,0,1)$ mode has $\bk_{101} = (2\pi/L_x,0,2\pi/L_z)$, while the $(2,0,2)$ mode has $\bk_{202} = (4\pi/L_x,0,4\pi/L_z)$. The theoretical growth rate of the fastest-growing fingering mode (see main text for detail) fits the observed exponential growth of the total kinetic energy at early times very well. The theoretical growth rate for the $(1,0,1)$ collective mode is $\Lambda_{101} = 0.0094 \pm 0.07i $, while $\Lambda_{202} = 0.02 \pm 0.08i$. Both compare quite well to the corresponding collective mode growth observed in the simulation. By $t = 300$, the growth of both collective modes seem to have saturated, presumably due to their mutual nonlinear interactions.}
\label{fig:collmodes}
\end{center}
\end{figure}

\begin{figure}[h]
\begin{center}
\includegraphics[width=\textwidth]{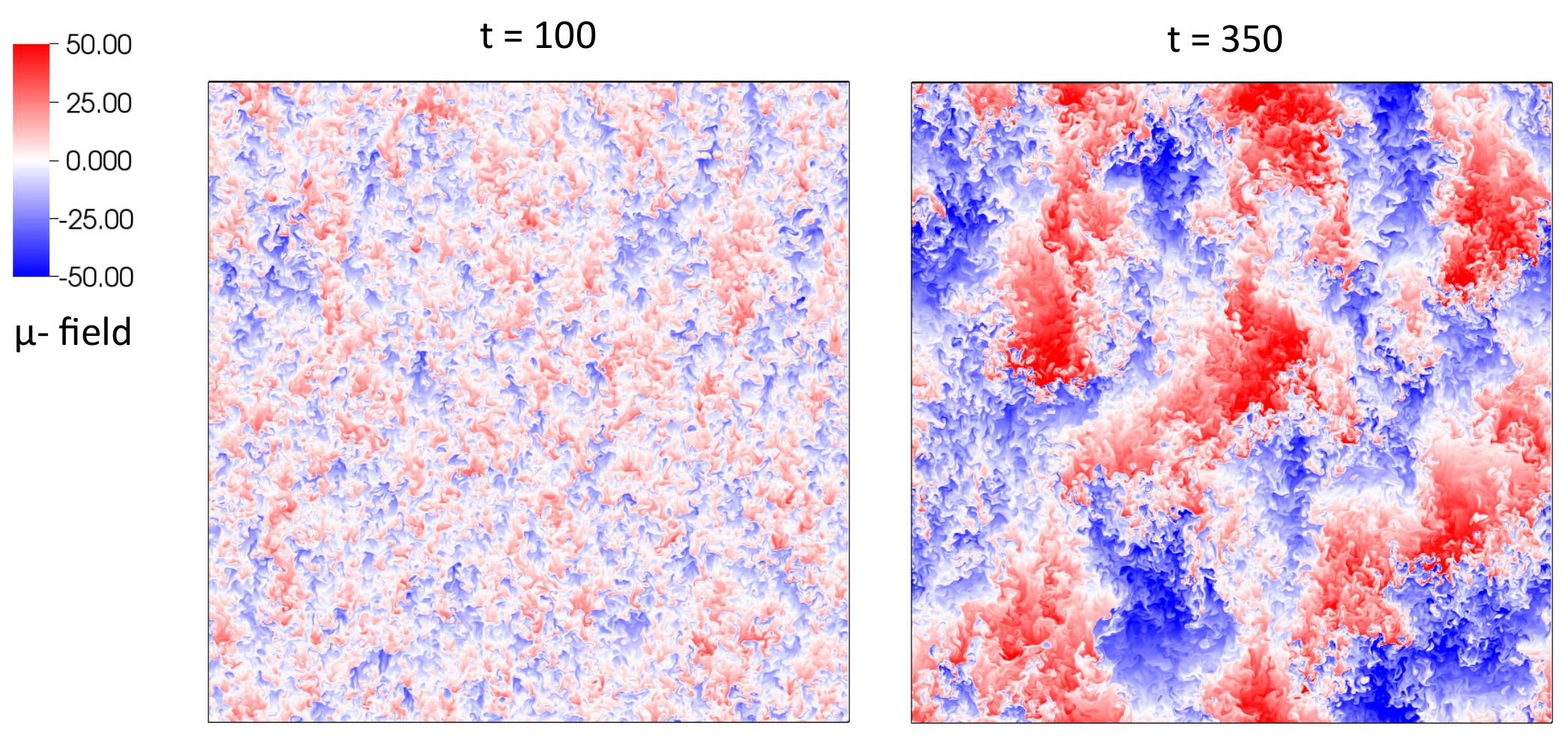}
\caption{Snapshots of the compositional field in our simulation at time $t=100$ (left) and $t = 350$ (right). At the early time, the fingering field is fairly homogeneous, but at the later time, a collective mode with two wavelengths in each direction (the $(2,0,2)$ mode) is clearly visible.}
\label{fig:Snaps}
\end{center}
\end{figure}

The simulation is initialized with small random fluctuations in the temperature field, which rapidly grow according to the dynamics of the fingering instability. This is illustrated in Figure \ref{fig:collmodes}, which shows the total kinetic energy in the domain as a function of time. As described by \citet{Brownal2013}, this initial linear instability then saturates as a result of the shear that develops between the fingers. This saturation can be seen in the first plateau observed shortly after $t = 90$. Figure \ref{fig:Snaps}a shows a snapshot of the compositional perturbations at $t = 100$, revealing a fairly homogeneous fingering field. Shortly thereafter, various large-scale gravity waves excited by the collective instability begin to grow. Figure \ref{fig:Snaps}b shows the same simulation as in Figure \ref{fig:Snaps}a, but at a later time when collective modes are clearly visible. 
Figure \ref{fig:collmodes} shows the kinetic energy in some of the largest-scale collective modes, and compares the observed growth rates with the corresponding ones predicted by the theory outlined in Section \ref{sec:collinst}. We find that our predictions are typically quite good, which confirms the assumptions made in Section \ref{sec:collinst}.

The collective instability appears to saturate when the kinetic energy in individual modes reaches a small fraction (a few percent) of the total kinetic energy. However, given that each mode is part of a mode family\footnote{a mode family being all modes of the kind $(\pm k_x,0,\pm k_z)$} whose members all grow at the same rate and saturate at the same amplitude, saturation occurs in fact when there is roughly equipartition between  the energy in all the fingering modes and the energy in all the collective modes. 

\begin{figure}[h]
\begin{center}
\includegraphics[width=\textwidth]{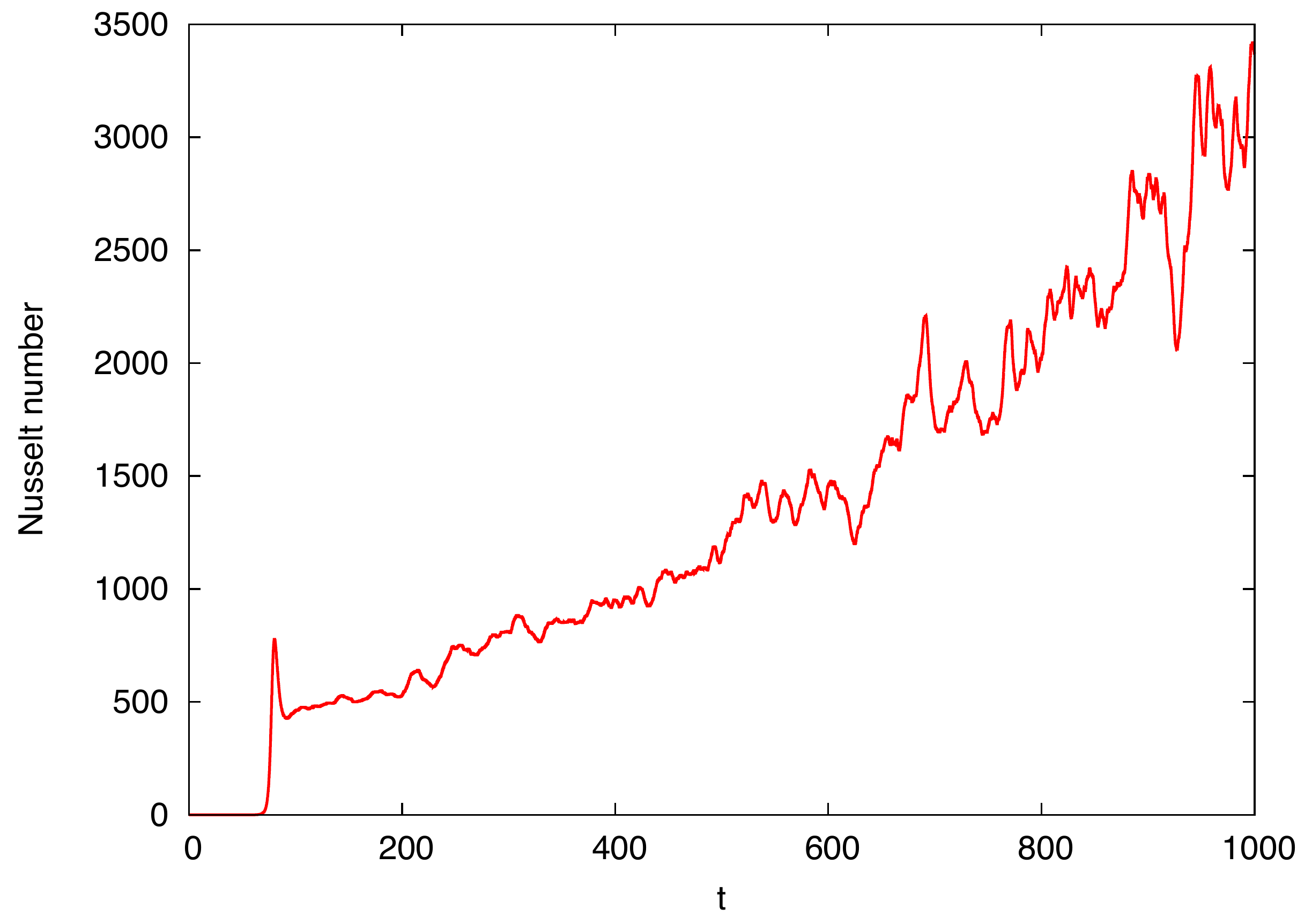}
\caption{Compositional Nusselt number ${\rm Nu}_\mu = D_\mu/ \kappa_\mu$. Two layers form in the simulation, around $t = 750$.}
\label{fig:Nusselt}
\end{center}
\end{figure}

\begin{figure}[h]
\begin{center}
\includegraphics[width=\textwidth]{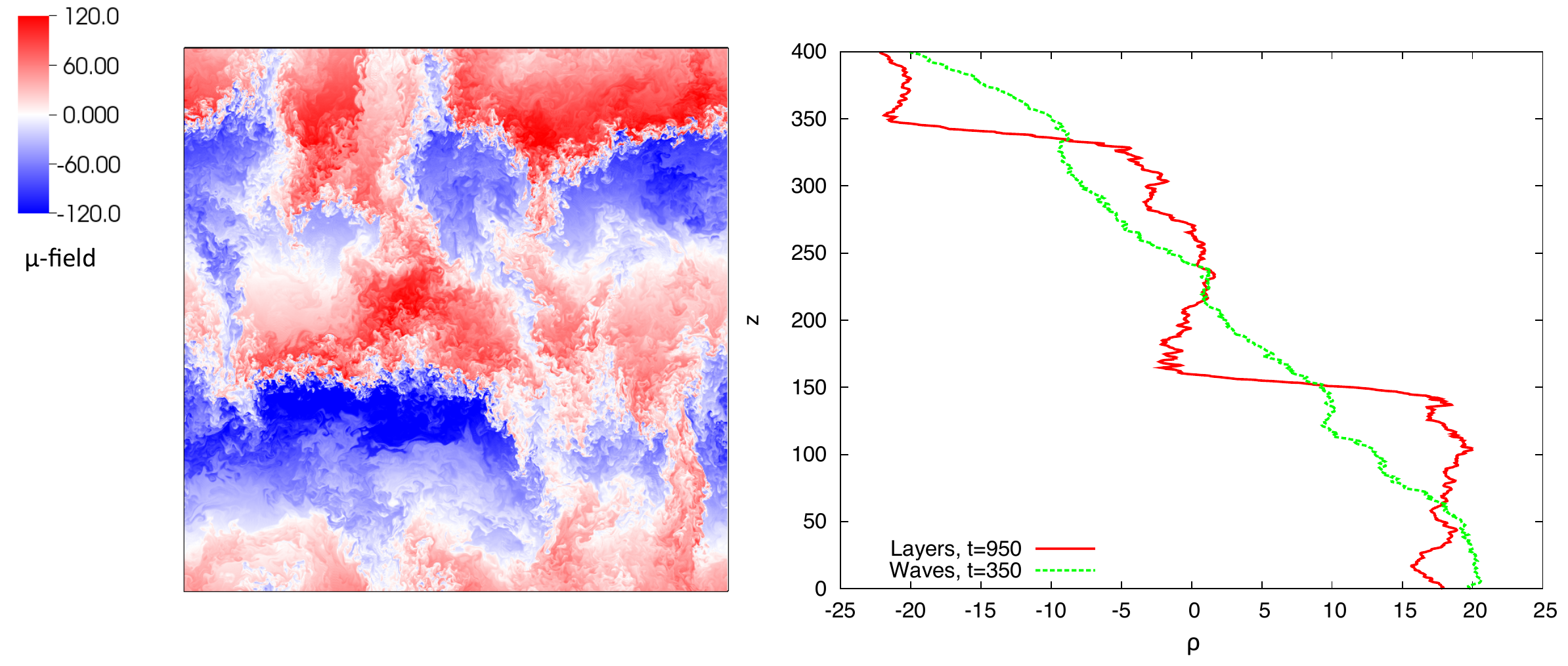}
\caption{Snapshot of the compositional field (left) and horizontally-averaged density profile (right) in our simulation at time $t=950$. Two convective layers with nearly uniform density, separated by thin and strongly stratified interfaces, are clearly visible. Also plotted for comparison is the density profile at $t = 350$, showing that the gravity waves themselves do not affect the mean density profile much.}
\label{fig:layers}
\end{center}
\end{figure}

As seen in Figure \ref{fig:Nusselt}, the growth of the waves causes a gradual increase in the global mixing rate. This can be quantified through the compositional Nusselt number, which is the ratio of the effective turbulent diffusivity $D_\mu$ to the microscopic diffusivity of the mean molecular weight:
\begin{equation}
{\rm Nu}_\mu = \frac{D_\mu}{\kappa_\mu}  \, .
\end{equation} 
Interestingly, we see that ${\rm Nu}_\mu$ continues to increase even when the wave amplitudes no longer do. This signals the onset of layered convection and confirms the results of \citet{Brownal2013}, who also found that their simulation transitioned into layers for the same set of parameters. Layered convection is characterized by a horizontally-averaged density profile that takes the form of a staircase, with region of near-uniform density (the convective layers) separated by strongly stable interfaces. We see in Figure \ref{fig:layers} that two layers form, with a spacing similar to the wavelength of the fastest-growing collective mode. This is consistent with the results of \citet{Traxleral2011b} in the oceanographic context, who also found that emerging thermo-compositional layers have the same wavelength as that of the fastest-growing collective modes in their simulations. By contrast with the high-Prandtl nature of oceanic fingering, the layers are not very robust, and are constantly pierced by strong updrafts and downdrafts that ultimately control the chemical transport rate in this system. By the end of the simulation (around $t = 1000$), we begin to see hints that the two layers may be merging as a result of these strong mixing events. In order to study the overall dynamics and general transport properties of layered convection in this regime, however, numerical simulations in domains of more even aspect ratio are needed. Indeed, at this point the quasi-two-dimensional nature of the simulation presented here becomes a hindrance as the convective eddies are severely restricted by the aspect ratio of the domain. Given the resolution required, this run would effectively have to be a $(1500)^3$ run, which is beyond the scope of this paper. 

It is worth noting that the mechanism by which thermo-compositional layers form in our simulation remains to be determined. Indeed, the standard layering transition caused by the $\gamma$-instability cannot occur at this parameter regime, as demonstrated by \citet{Traxleral2011}. An alternative possibility is that these layers are forming as the result of the nonlinear development of the collective modes, as originally discussed by \citet{stern2001sfu} in the oceanographic context. This scenario is much more difficult to confirm, since it can only be studied using DNS. Hence, whether thermo-compositional layering is an inevitable consequence of the collective instability, or whether it only occurs in a subset of parameter space remains to be determined. 

\section{Application throughout the HR diagram}
\label{sec:HRdiag}

We now look throughout the HR diagram to see whether any stars harbor regions of relatively high Prandtl number and diffusivity ratio. We focus on cases that are not fully convective, thereby ignoring very low-mass stars and stars on the Hayashi track. We split our study between stars on the main sequence, and post main sequence objects. 

\subsection{Main sequence stars}
\label{sec:MSstars}

The presence of inverse $\mu$-gradients is fairly rare in main sequence stars. This is because nuclear reactions, which tend to increase the mean molecular weight, are more efficient at higher temperatures, hence closer to the core. The gravitational settling of He with respect to H during the main sequence also usually contributes to the formation of a strong stable $\mu$-gradient. There are, however, three notable exceptions to this general statement. The first is the possibility of $^3$He burning, which is the only common reaction that decreases the mean molecular weight of the material. As first discussed by \citet{Ulrich1971} and \citet{Ulrich1972}, this effect can drive fingering convection, and can be important in RGB stars  for instance (see section \ref{sec:RGBstars}). The second exception is through the formation of thin element-accumulation layers near the stellar photosphere through the combination of radiative levitation and gravitational settling \citep{michaud70,charpinet97,richard01}. These layers can also be unstable to fingering convection, as first discussed by \citet{theado09} and more recently modeled by \citet{Zemskovaal2014}.

Finally, fingering convection can also be triggered by the accretion of higher-metallicity material on the surface, either via planetary infall \citep{vauclair2004mfa,Garaud11} or from a more evolved companion \citep{Ulrich1972,stancliffe2007cem}. For stars roughly below 1.5$M_\odot$, the accreted material is first rapidly mixed within their outer convection zone (which may deepen slightly as a result of the added metallicity). This creates an inverse $\mu$-gradient at the bottom of the convection zone, which can be unstable to fingering convection under the right circumstances. For stars above $1.5M_\odot$, which normally have a radiative envelope, the accreted material may at first  trigger the formation of a shallow outer convective layer. The latter rapidly mixes the added material, then disappears leaving behind a strong $\mu$-gradient, which can drive a fingering instability. The possibility of subsequently triggering the collective instability in the process has however never been investigated. 

We now look at stars in the mass range $0.7M_\odot$-$30M_\odot$, and determine their Prandtl number and diffusivity ratio.  Since we are only interested in an order of magnitude estimate, we approximate the thermal diffusivity by its radiative contribution:
\begin{equation}
\kappa_T =  \kappa_{\rm rad} = \frac{16 \sigma T^3}{3\kappa \rho^2 c_p} \, ,
\end{equation}
where $\sigma$ is the Stefan-Boltzmann constant, $T$ is the temperature, $\kappa$ is the opacity, $\rho$ is the density, and $c_p$ is the specific heat at constant pressure. We estimate the viscosity from the sum of its radiative and collisional contributions: 
\begin{equation}
\nu = \nu_{\rm rad} + \nu_{\rm coll} =  \frac{16\sigma T^4}{15c^2\kappa \rho^2} + \frac{0.406m_H^{1/2}(k_BT)^{5/2}}{e^4 C(\ln \Lambda_{\rm HH}) \rho}  \, ,
\end{equation}
where $c$ is the speed of light, $k_B$ is the Boltzmann constant, $e$ is the electron charge, $m_H$ is the mass of the hydrogen atom, and $\ln \Lambda_{ij}$ is the Coulomb logarithm for the collisions between element $i$ and element $j$:
\begin{equation}
\ln \Lambda_{ij} =  -19.26  - \frac{1}{2}\ln \rho + \frac{3}{2}\ln T - \frac{1}{2}\ln\left(\frac{X+3}{2}\right)  - \ln (Z_i Z_j) \, ,
\end{equation}
where $Z_i$ is the charge of element $i$, and $Z_j$ is the charge of element $j$. The expression for $\nu_{\rm coll}$ is from \citet{Spitzer1965}. The function $C(\ln \Lambda)$ is a correction to the Coulomb logarithm suggested by \citet{ProffittMichaud1993}, and is given by 
\begin{equation}
C(\ln \Lambda) = \frac{1}{1.2}  \ln \left(  \exp(1.2 \ln \Lambda) + 1 \right) \, .
\end{equation} 
For simplicity, when estimating the viscosity we use a fully ionized pure hydrogen gas, in which case $Z_i = Z_j = 1$ and $X = 1$. The correction to the Coulomb logarithm for a H-He mixture is of the order of a few percent, and is neglected here. These estimates for $\kappa_T$ and $\nu$ are valid as long as the stellar material is non-degenerate.

Finally, we use the diffusion coefficient given by \citet{ProffittMichaud1993}
\begin{equation}
\kappa_\mu =  \frac{15}{16 \rho \ln \Lambda_{\rm HHe}} \sqrt{\frac{2 m_H}{5 \pi}} \frac{(k_BT)^{5/2}}{e^4} \frac{3+X}{(1+X)(3+5X)(0.7+0.3X)} \, ,
\label{eq:kappamu}
\end{equation}
for the diffusion of He in a H gas (or vice versa). This coefficient overestimates the one for the diffusion of heavier elements in a H-He mixture\footnote{A very rough order-of-magnitude estimate of the diffusion coefficient of a trace species in a H-He mixture can be made by dividing $\kappa_\mu$ given in equation (\ref{eq:kappamu}) by $Z^2$, where $Z$ is the atomic charge of the species considered \citep{ProffittMichaud1993}. Using equation (\ref{eq:kappamu}) instead therefore overestimates the true diffusivity by a factor of a few to a few hundreds depending on $Z$. As we show later, $\kappa_\mu$ as calculated using (\ref{eq:kappamu}) is already too small to give rise to the collective instability. Accounting for the $Z$ correction can only reinforce our conclusion rather than change it.}, but the difference does not affect our estimate of where the collective instability may occur. 
            
\begin{figure}[h]
\begin{center}
\includegraphics[width=\textwidth]{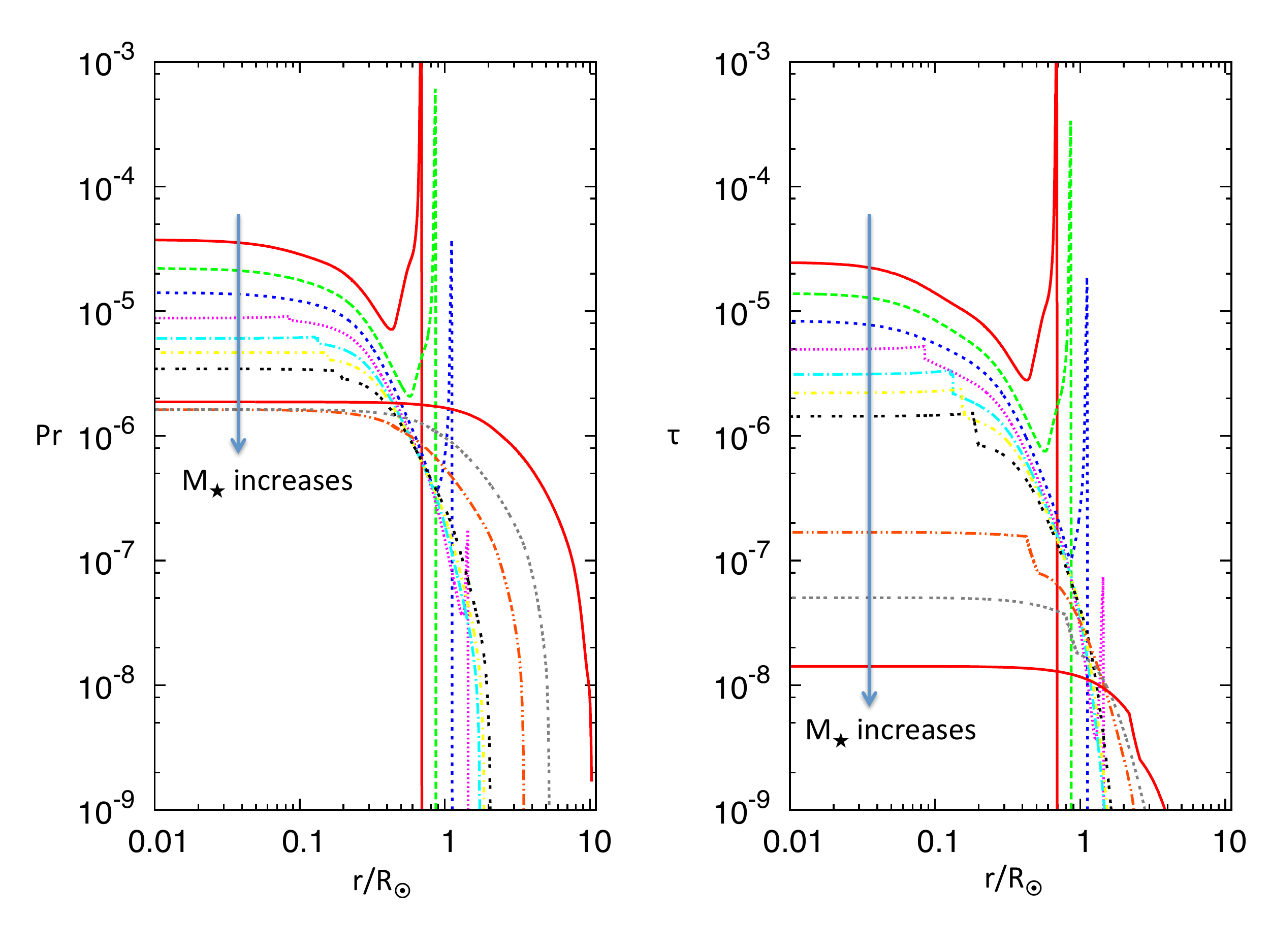}
\caption{Left: Prandtl number $\nu/\kappa_T$ as a function of radius in stars of various masses: from top to bottom following the arrow, $M_\star = 0.7M_\odot$, $0.9M_\odot$, $1.1M_\odot$, $1.3 M_\odot$, $1.5M_\odot$,  $1.7M_\odot$, $2M_\odot$, $5M_\odot$, $10M_\odot$ and $30M_\odot$. Right: Diffusivity ratio $\kappa_\mu/\kappa_T$ as a function of radius for the same stars. }
\label{fig:MSstars}
\end{center}
\end{figure}

The Prandtl number $\Pr = \nu/\kappa_T$, and the diffusivity ratio $\tau = \kappa_\mu/\kappa_T$ thus calculated are shown in Figure \ref{fig:MSstars} as a function of radius for stars of different masses. The stellar models used to calculate these ratios are obtained using the MESA stellar evolution code\footnote{Version 6794} \citep{Paxtonal2011,Paxtonal2013} to evolve each star from zero-age main sequence until their central mass fraction of hydrogen drops below 0.4. In each case, the star has initial solar composition, and the only mixing considered was due to standard convection. Note that while we have picked a particular stellar evolutionary stage at which to present this data, it is a good representation of both $\Pr$ and $\tau$ for these stars across the entire main sequence. 
We see that the diffusivity ratio is always smaller than the Prandtl number, by a factor of a few in the lower-mass stars (where $\nu$ is dominated by collisions), and by a much larger factor in high-mass stars (where it is dominated by radiation).  Generally-speaking, we also see that the only regions which contain material with a large Prandtl number are near the surface layers of stars of mass lower than $1.3M_\odot$, which corresponds precisely to the outer parts of their convective regions which cannot support fingering. We therefore conclude that the collective instability is unlikely to play any role in main sequence stars of any mass or age, and that the model proposed by \citet{Brownal2013} remains a good 
estimate for the rate of mixing by fingering convection wherever it occurs. 
 
\subsection{RGB stars}
\label{sec:RGBstars}

Fingering convection is more common after the main sequence turn-off, during off-center (shell-) burning phases. This is the case for instance in RGB stars, where $^3$He-burning on the outer edge of the H-burning shell creates a small, albeit significant, inverse $\mu$-gradient \citep{Eggleton06}. The effect of fingering convection in that region has been explored in depth by \citet{CharbonnelZahn07} and \citet{Denissenkov2010} for instance. Later on, fingering convection is also expected during the core He flash, as first discussed by \citet{Ulrich1972}. There, the off-center burning of He generates higher-$\mu$ material in a thin shell surrounding the core, and the lower part of this shell can become fingering-unstable. 

The core of RGB stars is mostly composed of pure fully ionized He, where the electrons are degenerate (or at least partially degenerate) while the nuclei remain non-degenerate. In this case, the thermal conductivity and shear viscosity are dominated by electron conduction, while the compositional diffusivity  remains dominated by collisional processes between the nuclei themselves. We use the formulae for the shear viscosity $\eta$ and thermal conductivity $k_T$ of strongly degenerate electrons given by \citet{Hubbard1966}, namely
\begin{eqnarray}
\eta = \frac{8}{135}  \frac{ \hbar^5 }{\pi^2 m_e^2 e^4 Z^2}  \left(\frac{4\pi \rho}{3Am_H}\right)^{5/3}  \kappa_F^8 H_\Gamma(\kappa_F) \, ,\nonumber \\
k_T = \frac{(2\pi \hbar)^3 k_B^2}{16 m_e^2 e^4 A m_H} G_\Gamma(\kappa_F) \rho T \, .
\end{eqnarray}
where $\hbar$ is the Planck constant, $A$ is the average atomic number of the nuclei, $m_e$ is the electron mass, and $\kappa_F$ is the dimensionless Fermi wavenumber $\kappa_F = (9\pi Z/4)^{1/3}$. The functions $H_\Gamma$ and $G_\Gamma$ are
\begin{equation}
G_\Gamma(x) = \frac{1}{\ln\left(1+\frac{4x^2}{3\Gamma}\right)} \mbox{  and  }  H_\Gamma(x) = \frac{G_\Gamma(x)}{2 - 2 G_\Gamma(x) + \frac{3\Gamma}{2x^2}} \, ,
\end{equation}
where, finally, 
\begin{equation}
\Gamma = \frac{Z^2 e^2}{k_BT} \left(\frac{ 4\pi \rho}{3 A m_H} \right)^{1/3} \, .
\end{equation}
These expressions are only valid in a strongly degenerate limit, in which the degeneracy parameter $C_2$ is small, with 
\begin{equation}
C_2 = \frac{1}{6} \left( \frac{m_e k_B Z^{1/3}}{\hbar^2} \right)^{2} \left(\frac{Am_H}{2 Z \rho}\right)^{4/3} T^2  \, .
\end{equation}
The thermal diffusivity and kinematic viscosity due to electron conduction in this highly degenerate limit are then given by $\nu_e = \eta /\rho$ and $\kappa_{e} = k_T /\rho c_p$, as usual. For the purpose of the following calculation we use $A = 4$ and $Z = 2$ since the core is mostly composed of He.

For simplicity, because we are again only interested in an order-of-magnitude estimate of the Prandtl number and of the diffusivity ratio, we compute the total thermal diffusivity and kinematic viscosity by adding the degenerate electron contribution to the usual non-degenerate terms, as 
\begin{eqnarray}
\kappa_T =  \kappa_{\rm rad} + \kappa_{e} \, , \nonumber \\
\nu = \nu_{\rm  rad} + \nu_{\rm coll} + \nu_e \, .
\end{eqnarray}
The non-degenerate nuclei contributions typically dominate in the star's envelope, while the electron conduction contribution dominates in the core. Simply adding the two contributions is adequate either in the highly-degenerate or in the non-degenerate limits, the relative error committed being largest (and of order unity) in the intermediate region of weak degeneracy where $C_2 \sim 1$. Finally, since the nuclei are not degenerate, we use the same formula for $\kappa_\mu$ as the one given in equation (\ref{eq:kappamu}), with $\ln \Lambda_{\rm HHe}$ replaced by $\ln \Lambda_{\rm HeC}$.

\begin{figure}[h]
\begin{center}
\includegraphics[width=\textwidth]{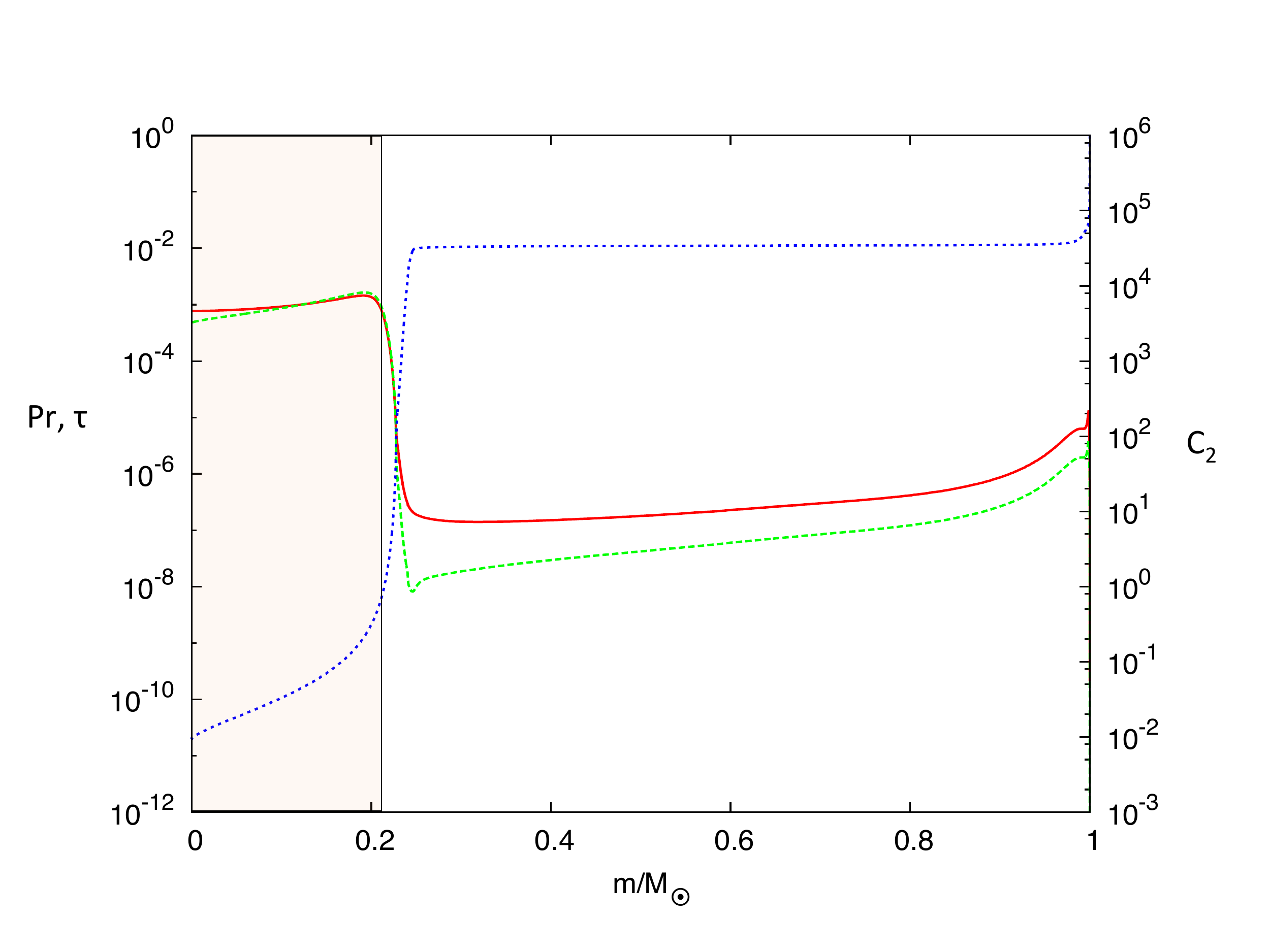}
\caption{Prandtl number (solid red line), diffusivity ratio (green dashed line) and degeneracy parameter $C_2$ (see text for detail, blue dotted line) for a $1\,M_\odot$ star just before the luminosity bump at $t=11.47$Gyr. The left axis is for $\Pr$ and $\tau$, while the right axis is for $C_2$. The shaded area mark the region of relatively high degeneracy ($C_2 < 1$) which is bounded from above by the H-burning shell. The values of $\Pr$ and $\tau$ are least accurate around $C_2 = 1$. }
\label{fig:RGBstars}
\end{center}
\end{figure}

\begin{figure}[h]
\begin{center}
\includegraphics[width=\textwidth]{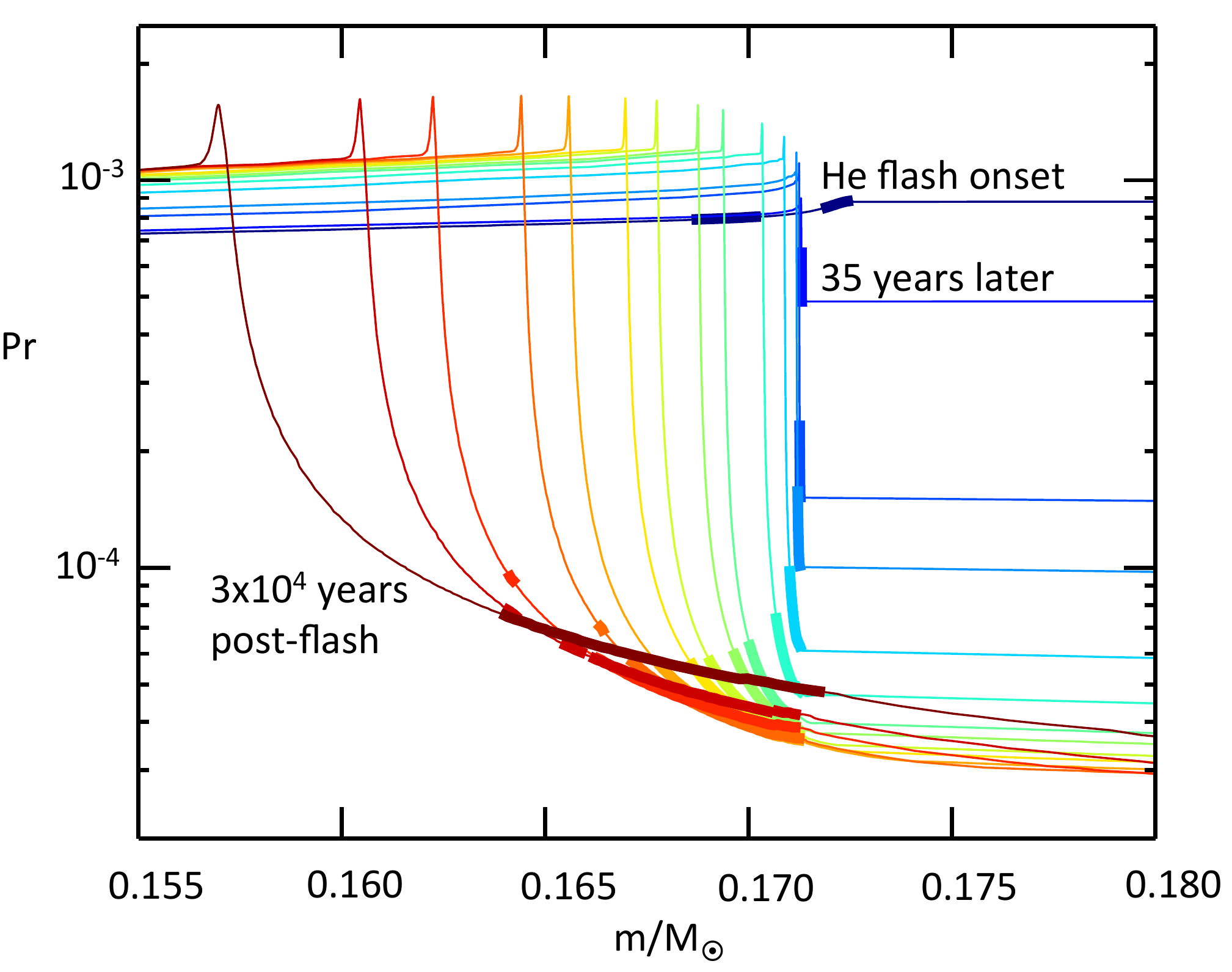}
\caption{Variation of the Prandtl number with mass coordinate as a function of time from the onset of the first core He flash (at $t = 11.58$Gyr) for a $1M_\odot$ star. The entire core itself at that time spans the interval $m \in [0,0.45 M_\odot]$; only the region near the He-burning shell is shown. The line color is used as a proxy for time (from dark blue at the earliest time, to dark red at the latest time). Regions that are unstable to fingering convection are shown as a thick line in each profile. At the onset of the flash, and for a very brief period thereafter, it is possible to find fingering regions with high Prandtl number. However, the flash partially lifts the electron degeneracy, which causes the Prandtl number to decrease rapidly. }
\label{fig:Heflash}
\end{center}
\end{figure}

The profiles of the Prandtl number $\Pr$ and diffusivity ratio $\tau$ in a $1\,M_\odot$ stellar model are shown in Figures \ref{fig:RGBstars} and \ref{fig:Heflash} at several points on the RGB.  Figure \ref{fig:RGBstars} shows a profile taken just before the luminosity bump, which corresponds to the onset of fingering convection in low-mass stars \citep{CharbonnelZahn07}. Figure \ref{fig:Heflash} shows a number of profiles taken during the first core He flash at the tip of the RGB. These models were obtained using the MESA stellar evolution code\footnote{Version 7385} to evolve a 1$M_\odot$ nonrotating star with metallicity $Z=0.02$ starting from the pre-main sequence, including atomic diffusion and gravitational settling. 

In both Figures, we see that the Prandtl number and the diffusivity ratio are much larger in the core where microscopic transport is dominated by electron conduction. By contrast they are quite small in the non-degenerate envelopes, as it is the case for main sequence stars. This suggests that the only potential place where the collective instability could be at play is in regions of strong electron degeneracy. Having a region of relatively high Prandtl number and diffusivity ratio is not enough, however, to guarantee the existence of the collective instability. This region must also be the subject of active fingering convection, and the density ratio must be sufficiently low as discussed in Section \ref{sec:collinst}.

For stars ascending the RGB, the inverse $\mu$-gradient begins on the outskirt of the H-burning shell, distinctively outside the core. For the 1$M_\odot$ star shown in Figure \ref{fig:RGBstars}, this corresponds to the region between $m/M_\odot = 0.24$ and the bottom of the convection zone. We see that the Prandtl number is lower than $10^{-4}$. Furthermore, the edge of the core itself is the seat of a very strong {\it stabilizing} $\mu$-gradient, preventing any of the turbulence induced by the fingering instability from penetrating into the degenerate region. We therefore confirm the results of \citet{DenissenkovMerryfield2011}, namely that the collective instability is unlikely to play a role in these stars. 

The case of stars undergoing core He flash is {\it a priori} more promising: the initial flash takes place well-within the region of strong electron degeneracy where the Prandtl number is roughly $10^{-3}$, a value for which the collective instability is indeed active.  The off-center He flash convection zone converts a few percent of its $^4$He by mass into $^{12}$C via the triple-alpha process, leaving an inverse $\mu$-gradient at its bottom boundary.  The stabilizing influence of the background temperature gradient is very marginal, since the core itself is close to being isothermal.  Thus one may also expect to find regions of very low density ratio $R_0$ (also necessary for the development of the collective instability).  We note, however, that the first He flash partially removes the electron degeneracy while the burning proceeds (see Figure \ref{fig:Heflash}), so that one can only find a fingering region at relatively high Prandtl number for a very short time. Subsequent flashes take place in weakly degenerate conditions with $\Pr$ and $\tau$ of the order of $10^{-4}$, which is too low for the collective instability to exist. As a result, the collective instability can at best only be active in a very short timeframe, and is therefore unlikely to play a significant role in stars undergoing core He flash. 

\subsection{White dwarfs}
\label{sec:WDstars}

As for main sequence stars, fingering convection is not typically thought to occur spontaneously in white dwarfs. However, the accretion of high-$\mu$ material on their surface, from planetary debris or from winds from stellar companions, could trigger the double-diffusive instability, as recently discussed by \citet{Dealal2013}. 

\begin{figure}[h]
\begin{center}
\includegraphics[width=\textwidth]{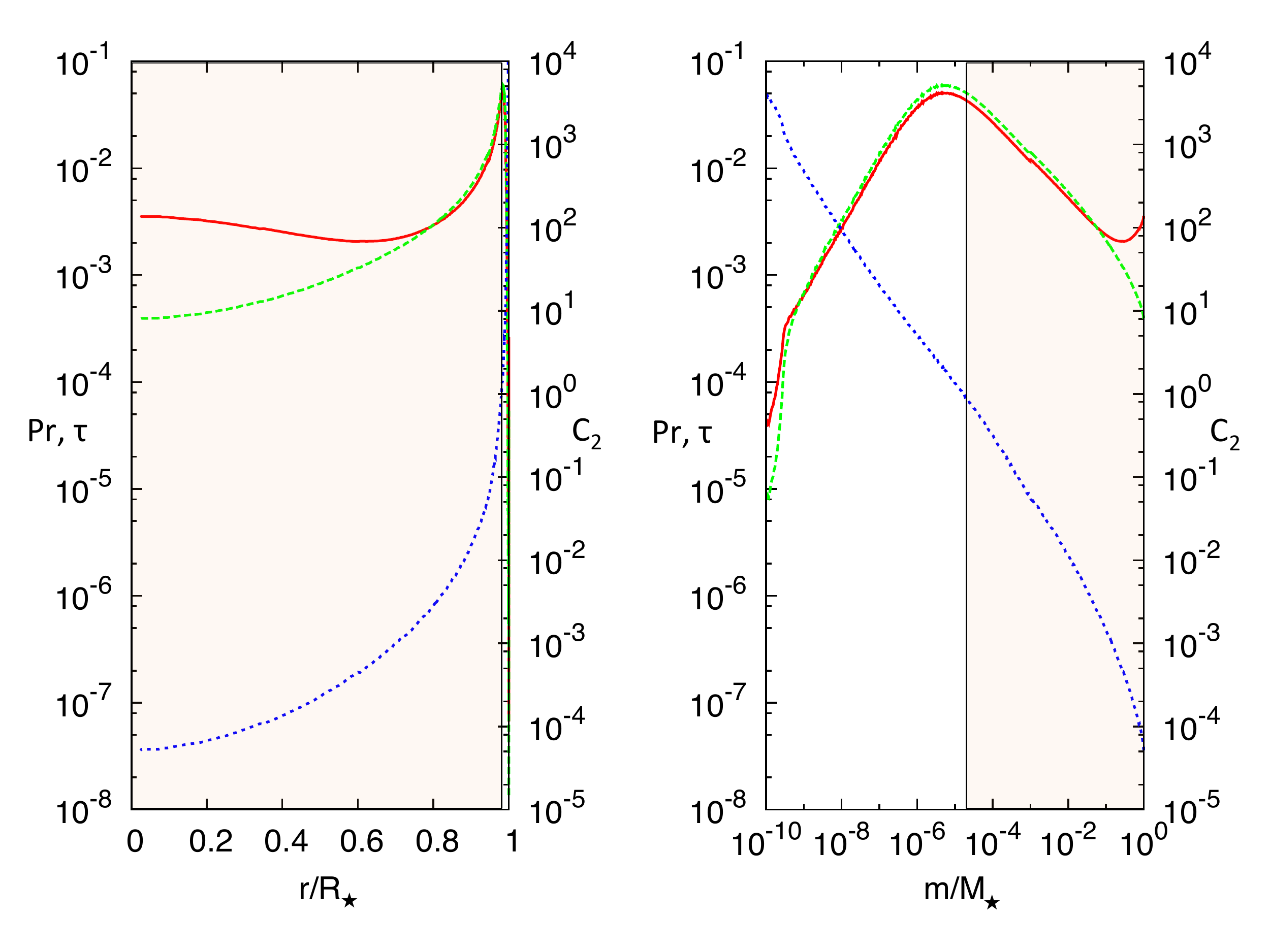}
\caption{Prandtl number (solid red line), diffusivity ratio (green dashed line) and degeneracy parameter $C_2$ (see text for detail, blue dotted line) in the white dwarf model described in Section \ref{sec:WDstars}. In both panels, the left axis is for $\Pr$ and $\tau$, while the right axis is for $C_2$. The shaded and white areas mark the region of relatively high (resp. low) degeneracy ($C_2 < 1$ for the shaded areas). The values of $\Pr$ and $\tau$ are least accurate around $C_2 = 1$.  Left: Profiles as a function of radius. Right: Profiles as a function of mass coordinate starting from the surface at $m=0$. }
\label{fig:WDstars}
\end{center}
\end{figure}

Since electron degeneracy is high in a white dwarf except very close to the surface, we expect the Prandtl number and the diffusivity ratio to be relatively high as well, as in the cores of RGB stars. This is indeed the case, as shown in Figure \ref{fig:WDstars}. For the purpose of illustration, we present a DA white dwarf model with effective temperature $T_{\rm eff} = 11150$K, kindly provided by G. Vauclair and S. Vauclair. This model was computed in the same manner as the DA white dwarf models of \citet{Dealal2013}, and its parameters were chosen to be representative of G29-38, the prototype DA white dwarf with a known debris disk and polluted atmosphere \citep{Koesteral97}. Its mass is $M_\star = 0.59 M_\odot$, and its radius is $R_\star = 0.0138R_\odot$. The Prandtl number and the diffusivity ratio are calculated as in Section \ref{sec:RGBstars} (except that we use $A = 12$ and $Z = 6$ in the core of the star). For the purpose of clarity, we also show them as functions of the mass coordinate starting from the surface (i.e. with $m = 0$ corresponding to the surface, and $m = M_\star$ corresponding to the core). Both ratios are quite small in the non-degenerate atmosphere (down to $m/M_\star \sim 10^{-6}$), then increase with depth to be of the order of a few times $10^{-3}$ in the rest of the star, suggesting that the collective instability might possibly be excited there. 

As with RGB stars, having a relatively high Prandtl number and diffusivity ratio is not enough to guarantee the presence of the collective instability. We must first determine whether the fingering convection excited by material accretion may penetrate down to $m/M_\star \sim 10^{-6}$ and whether the density ratio at this level is indeed low. In order for the first condition to be satisfied, the mean molecular weight of the accreting material must be relatively large, at least larger than that of the non-degenerate envelope, and ideally larger than that of the core. This is not unlikely if its origin is from dust and planetary remnants. Whether a sufficiently low density ratio can also be created in the process remains to be determined with detailed calculations along the lines of those presented by \citet{Dealal2013} for DA white dwarfs. This will be the subject of a follow-up investigation.


 \section{Conclusion and observational prospects}
 \label{sec:ccl}

In this paper, we have shown through theory and numerical experiments that fingering (thermohaline) convection in stars can in principle drive large-scale gravity waves through a mean-field instability called the ``collective instability''. Note that by large-scale we imply waves whose typical wavelengths are much larger than that of individual fingering modes, although from a global stellar point of view they remain quite small-scale (see below for more on this topic). We have performed a comprehensive exploration of parameter space to determine under which conditions the collective instability is excited, and concluded that it is limited to stellar regions in which the Prandtl number is relatively large (at least as large as $10^{-3}$). This only happens when electron degeneracy is high and dominates thermal conduction and viscous dissipation, such as in the cores of post main sequence stars, and white dwarfs. Compounded with the fact that the system must first and foremost be undergoing active fingering convection, this only leaves two possibilities for plausible scenarios where the collective instability could play a role: RGB stars undergoing core He flashes (albeit only for a very short time after the onset of the first flash) and white dwarfs undergoing accretion of high-metallicity material from planetary debris or companion stellar wind. 

The excitation of the collective instability in these examples could have two different consequences. First, it is a natural source of gravity waves -- one could therefore wonder whether they may be directly observable through asteroseismology or not. Second, it causes a significant increase in the mixing rate (especially if convective layering occurs), which could in turn affect the evolution of the star, its internal structure and  its surface abundances. 

As discussed in Section \ref{sec:RGBstars}, the collective instability is only likely to be relevant in RGB stars in the first few tens of years after the onset of core He flash. The effect of enhanced mixing by fingering convection alone has (to our knowledge) never been investigated in detail, so we can only speculate on the added effects of mixing by collective modes. Both processes would likely act to erode compositional gradients established at the lower boundary of He flash convection regions. Whether this might have any observable signature, however short-lived, is unclear. Like all red giants, these stars exhibit ``mixed modes" which behave like acoustic modes in the convective envelope and like gravity modes in the radiative core \citep{Mosseral2011}. Their frequencies generally depend on the buoyancy frequency in the core, and are thus sensitive to composition gradients there. However, even without additional mixing, the magnitude of $N_\mu$ (the component of the buoyancy frequency associated with $\mu$-gradients) established in He flash regions is much less than that at the outer boundary of the He core. The latter largely dominates the mode spectrum and overwhelms any asteroseismic signature coming from the He flash regions themselves \citep{Hual09}. Since extra mixing can only reduce the small composition gradients resulting from He flashes, it would be difficult to deduce the presence of the collective instability from an observed mode spectrum. Finally, since $\mu$-inversions in the degenerate core only appear \emph{below} He flash convection regions, the collective instability cannot modify surface abundances during this phase. 

The possibility of exciting gravity waves through the collective instability in white dwarfs is potentially more interesting on the other hand for two reasons. First, and as studied by \citet{Dealal2013}, fingering convection alone has a significant effect on the redistribution of high-$\mu$ material accreted onto DA white dwarfs, and must be taken into account when estimating the accretion rate from observations of the surface chemical abundances. The additional mixing caused by the collective instability would go further in the same direction, and must therefore also be taken into account. Second, these waves could potentially be observable if they reach sufficiently large amplitudes and horizontal scales. For $\Pr \sim \tau \sim 10^{-3}$ and reasonably small density ratios, the horizontal and vertical wavelengths of the fastest-growing collective modes are of the order of $300d$ (see Figure \ref{fig:collmodes}), which in dimensional terms correspond to about 30 meters. While this is too small to be observable, we note that larger-scale modes also grow, albeit at smaller rates (see Figure \ref{fig:flowerplots}). Furthermore, the nonlinear interactions between growing modes can also excite very large-scale ones even if the latter would otherwise intrinsically decay. In other words, it is not unlikely that global-scale waves could be present as well. The question of the saturation amplitude of the collective instability, however, is much more complicated, and remains to be addressed. Nevertheless, these findings raise an interesting prospect, namely that of finding polluted white dwarf pulsators outside of the standard instability strips \citep[e.g. see][for a review on white dwarf pulsators]{WingetKepler08}. 

Finally, we have confirmed the findings of \citet{Brownal2013} concerning the spontaneous formation of thermo-compositional layers as a possible nonlinear progression of the collective instability. Why these layers form and under which conditions one may expect to find them remains to be determined. However, this could mean that the ultimate signature of the collective instability may not be through the excited waves, but through the presence of thermo-compositional layering. 
  
\acknowledgements

The authors thank G\'erard and Sylvie Vauclair for providing the white dwarf model presented in Section 4, and for their help with improving the clarity of the manuscript. They also thank W. Hubbard for his help. This work is funded by AST-1412951. J. B. is supported by the Department of Energy Office of Science Graduate Fellowship Program (DOE SCGF), made possible in part by  the American Recovery and Reinvestment Act of 2009, administered by ORISE-ORAU under contract no. DE-AC05-06OR23100.  K. M. was partly supported by AST-0847477 and AST-1211394. The numerical simulations presented here were done with the PADDI code kindly provided by S. Stellmach, and were run on the Hyades supercomputer at UCSC, which was purchased using an NSF-MRI grant.


\end{document}